\documentclass[a4paper,12pt]{article}
\usepackage{epsfig,amssymb,amsmath}
\usepackage{graphicx}
\usepackage{amsfonts}

\begin{document}
%Revised version to deal with referee requests

\newcommand{\be}{\begin{equation}}
\newcommand{\ee}{\end{equation}}
\newcommand{\bea}{\begin{eqnarray}}
\newcommand{\eea}{\end{eqnarray}}
\newcommand{\bean}{\begin{eqnarray*}}
\newcommand{\eean}{\end{eqnarray*}}
\newcommand{\nn}{\nonumber}
\font\upright=cmu10 scaled\magstep1
\font\sans=cmss12
\newcommand{\ssf}{\sans}
\newcommand{\stroke}{\vrule height8pt width0.4pt depth-0.1pt}
\newcommand{\C}{\mathbb{C}}
\newcommand{\CP}{\mathbb{CP}}
\newcommand{\Z}{\mathbb{Z}}
\newcommand{\half}{\frac{1}{2}}
\newcommand{\quart}{\frac{1}{4}}
\newcommand{\bphi}{\bar{\phi}}
\newcommand{\bPhi}{\bar{\Phi}}
\newcommand{\bz}{\bar{z}}
\newcommand{\bZ}{\bar{Z}}
\newcommand{\pr}{\partial}
\newcommand{\bm}{\boldmath}
\newcommand{\I}{{\cal I}} 
\newcommand{\Lag}{{\mathcal L}}
\newcommand{\M}{{\cal M}}
\newcommand{\N}{{\cal N}}
\newcommand{\V}{{\cal V}}
\newcommand{\e}{\varepsilon}
\newcommand{\g}{\gamma}
\newcommand{\Tr}{{\rm Tr}}

\thispagestyle{empty}
\vskip 3em
\begin{center}
{{\bf \LARGE A $\CP^2$ SMEFT}} 
\\[15mm]

{\bf \large N.~S. Manton\footnote{email: N.S.Manton@damtp.cam.ac.uk}} \\[20pt]

\vskip 1em
{\it 
Department of Applied Mathematics and Theoretical Physics,\\
University of Cambridge, \\
Wilberforce Road, Cambridge CB3 0WA, U.K.}
\vspace{12mm}

%%%%%%%%%%%%%%%%%%%%%%%%%%%%%%%%%%%%%%%%%%%%%%%%%%
\abstract
{An extension of the Standard Model is proposed, where the Higgs field
is valued in the complex projective plane $\CP^2$, rather than $\C^2$.
Its geometry is consistent with $U(2) \simeq (SU(2) \times U(1))/
\Z_2$ electroweak gauge symmetry. The leading terms in the Lagrangian,
beyond those of the Standard Model, are much more tightly constrained
than in general SMEFTs, and the custodial $SO(4)$ symmetry of the
Higgs sector is mildly broken. The predicted tree-level deviations
from Standard Model phenomenology depend on a single large mass
parameter $M$. Current experimental data imply that $M$ is a few TeV
or larger. 
}
%%%%%%%%%%%%%%%%%%%%%%%%%%%%%%%%%%%%%%%%%%%%%%%%%%

\end{center}

\vskip 150pt
%\leftline{Keywords: }
\vskip 1em

\vfill
\newpage
\setcounter{page}{1}
\renewcommand{\thefootnote}{\arabic{footnote}}

%%%%%%%%%%%%%%%%%%%%%%%%%%%%%%%%%%%%%%%%%%%%%%%%%%%%%%%%%%%%%%%%
\section{Introduction} 
%%%%%%%%%%%%%%%%%%%%%%%%%%%%%%%%%%%%%%%%%%%%%%%%%%%%%%%%%%%%%%%%

We propose here an extension of the Standard Model (SM) \cite{PS} in the
spirit of a SMEFT (Standard Model Effective Field Theory) \cite{Fal}.
Its perturbative field expansion has terms of all orders, but the dominant
non-renormalisable terms are of mass dimension 6. The key idea is to
exploit the geometry of the complex projective plane $\CP^2$ as the
manifold where the Higgs field takes its values. The Riemannian metric
on $\CP^2$ is assumed to be the Fubini--Study metric, the standard maximally
symmetric metric on $\CP^2$. In our model, its isometry group $SU(3)$
is present in the background, but explicitly broken by choosing a point on
$\CP^2$ and retaining only the $U(2)$ isotropy group of this point
as the gauge symmetry of the Lagrangian. In other words, our model is
a $\CP^2$ sigma-model supplemented by a Higgs potential, where the
$U(2)$ action is gauged.

Locally, near the chosen point of $\CP^2$, the
geometry resembles that of flat $\C^2$ near the origin, and there is one
complex-doublet Higgs field. Our $\CP^2$ model is therefore close to the
Standard Model, provided the geometric scale factor of the $\CP^2$ manifold
is large and the Higgs potential is chosen so that the Higgs vacuum manifold
is small on this scale. In units of the Higgs vacuum expectation value
(vev), the model therefore has a single large mass parameter $M$ for
Beyond the Standard Model (BSM) physics. 

In the electroweak sector of the Standard Model, including its gauge fields and
fermions, the symmetry group is $U(2) \simeq (SU(2) \times U(1))/ \Z_2$.
However, the usual Higgs potential has a larger, custodial $SO(4)$ symmetry,
and the Higgs vacuum manifold is a round 3-sphere. This custodial symmetry
is not exploited here, so our model has a different geometric structure
than models where it is retained \cite{AJM,AW}. However, the generic orbits
of $U(2)$ on $\CP^2$ are still three-dimensional; they are a family of
squashed 3-spheres. One orbit is simply a point, and in the
neighbourhood of this, the orbits are very close to being round
3-spheres. Therefore the custodial symmetry is approximately
retained, and its important consequence for the ratio of the
$W$-boson and $Z$-boson masses remains approximately valid. It was
observed in ref.\cite{AJM} that $SO(4)$ symmetry, broken to $U(2)$,
leads to squashed 3-spheres, but the connection to $\CP^2$ geometry
was not mentioned.

It is well known that there is no inconsistency at present in assuming
that the SM gauge group is $(SU(3) \times SU(2) \times U(1))/ \Z_6$.
The quantum numbers of the Higgs field and the known fermion multiplets
(quarks and leptons) lie in representations that are consistent with the
$\Z_6$ quotient, and there are no local or global anomalies \cite{DGL}.
In the analysis that follows, we will ignore the QCD interactions of
quarks and gluons. These sectors of the Standard Model Lagrangian are
unaffected by changes to the Higgs field geometry. We will discuss only the
electroweak sector, including the Higgs field, and the quark and
lepton Yukawa couplings to the Higgs field. The relevant gauge group
is therefore the electroweak $U(2)$ subgroup of the SM gauge group,
mentioned above. We will also ignore
fermion generation mixing, and simplify the neutrino physics to the
version that was considered in the 1960s and '70s, where there is
just a left-handed neutrino field; our model has no novel insights
concerning these challenging topics.
 
The Lagrangian has terms involving: (i) Gauge-covariant
derivatives of the Higgs field, (ii) a Higgs potential, (iii) Gauge
field tensors, (iv) Dirac (kinetic) terms for fermions, (v) Yukawa
terms coupling the Higgs field and fermions. We will discuss these
in turn, and then explore some phenomenological consequences, in
particular, for the $W$-boson to $Z$-boson mass ratio, for the
couplings of the $W$- and $Z$-bosons to the physical Higgs
boson\footnote{The earlier published version, JHEP 04 (2025) 180,
did not include this topic.}, and for the Yukawa couplings of the Higgs
boson. These will lead to constraints on the BSM mass parameter $M$.

Our presentation begins with a review of the Fubini--Study metric on
$\CP^2$.

%%%%%%%%%%%%%%%%%%%%%%%%%%%%%%%%%%%%%%%%%%%%%%%%%%%%%%%%%%%%%%%%%%%%%%%%%
\section{$\CP^2$ Geometry}
%%%%%%%%%%%%%%%%%%%%%%%%%%%%%%%%%%%%%%%%%%%%%%%%%%%%%%%%%%%%%%%%%%%%%%%%%

The complex projective plane $\CP^2$ is obtained from flat complex
3-space, $\C^3$, by quotienting out the action of scaling by a complex
number. Geometrically, $\CP^2$ is the manifold of complex
lines\footnote{We use the mathematician's notion of complex
line $\C$ to denote what is more commonly called the complex plane.}
through the origin in $\C^3$. 

We denote a non-zero point in $\C^3$ by the (column) vector $Z = (Z_0, Z_1,
Z_2)$ and the scaling action is $Z \to \kappa Z$ with $\kappa \in \C
- \{0\}$. It is convenient
to use the inhomogeneous coordinates $Z = (1, z_1, z_2)$ on
$\CP^2$. However, this misses the part of $\CP^2$ ``at infinity''. The
missing part is the manifold of points $Z = (0, Z_1, Z_2)$ modulo
scaling by $\kappa$, which is a copy of the complex projective
line $\CP^1$, and geometrically a 2-sphere. Most of our calculations
will be in the neighbourhood of $Z = (1, 0, 0)$ so the use of
inhomogeneous coordinates causes no difficulty.

The hermitian conjugate (row) vector $\bZ = (Z_0^*, Z_1^*, Z_2^*)$
will play an important role in what follows. In inhomogeneous
coordinates, the hermitian conjugate of $Z$ is $\bZ = (1, z_1^*,
z_2^*)$.

$\C^3$ has the Euclidean metric $ds^2 = d\bZ dZ$, whose
isometry group, fixing the origin, is $SO(6)$. The
natural metric on $\CP^2$ is derived from this by removing from
$dZ$ its projection parallel to $Z$, and then normalising so that
rescaling by $\kappa$ has no effect. The result is the Fubini--Study
metric \cite{FS}
\be
ds^2 = \frac {(\bZ Z) d\bZ dZ - (d\bZ Z) (\bZ dZ)}{(\bZ Z)^2} \,.
\label{FShom}
\ee
This is in dimensionless form; a scale factor will be introduced with
the Higgs field. Note that if $dZ = \mu Z$ with $\mu \in \C$, i.e.
$dZ$ parallel to $Z$, then $ds^2 = 0$.

The Fubini--Study metric is K\"ahler and of constant positive
curvature. Because of the projection mentioned above, it depends in
an essential way on the complex character of the coordinates. The
isometry group is therefore not $SO(6)$ but its subgroup $SU(3)$.
In homogeneous coordinates, the symmetry group action is simply
$Z \to UZ$, with $U \in SU(3)$. $SU(3)$ acts transitively on $\CP^2$,
and the isotropy group of any given point is $U(2)$. As a manifold,
therefore, $\CP^2 = SU(3)/U(2)$. We can check the isotropy group for
the point $Z = (1,0,0)$, as follows. The matrices $U \in SU(3)$
that preserve $Z$ are of the form
\be
U = \begin{pmatrix}
    e^{-2i\gamma} & 0 \\ 0 &  e^{i\gamma} u  
    \end{pmatrix} \,,
\ee
with $u \in SU(2)$. Naively, this sends $(1,0,0)$ to $(e^{-2i\gamma}, 0, 0)$,
but this is equivalent to $(1,0,0)$ in the projective space.
$e^{i\gamma} u$ is a unitary $2 \times 2$ matrix, and apparently an element of
$SU(2) \times U(1)$, but it is unchanged if $\gamma \to \gamma + \pi$
and $u \to -u$, so the isotropy group is indeed $U(2)$.

In our model, the point $(1,0,0) \in \CP^2$ plays a preferred role, and its
isotropy group $U(2)$ will be identified with the electroweak gauge
group. The generic orbits of $U(2)$ on $\CP^2$ are 3-spheres, as will
be clarified in the next section.

Later, we will be using the projections of a generic vector $\Psi \in
\C^3$, defined at location $Z \in \C^3$, on to the line through the
origin and $Z$ and on to its complement, which is a copy of the tangent space to
$\CP^2$. These projections are respectively
\be
\Psi \to P\Psi = \frac{Z(\bZ \Psi)}{\bZ Z}
\ee
and
\be
\Psi \to (1-P)\Psi = \frac{(\bZ Z)\Psi - Z(\bZ \Psi)}{\bZ Z} \,.
\ee
Because these projections are unaffected by rescaling $Z$ by
$\kappa$, they are defined at location $Z$ on $\CP^2$. More precisely,
over $\CP^2$ there is its 1-dimensional tautological bundle, whose fibre
at a point $Z \in \CP^2$ is the line in $\C^3$ through the origin and $Z$,
and its 2-dimensional tangent bundle. The 3-dimensional, trivial $\C^3$
bundle over $\CP^2$ is the direct sum of these, so an element $\Psi$ in the
trivial $\C^3$ bundle is the sum of its projections on to the
tautological and tangent bundles.

Note that $\bZ \Psi$ and $\bZ Z$ are $SU(3)$-invariant. Therefore, both
$P\Psi$ and $(1-P)\Psi$ transform in the same way as $\Psi$ under
$SU(3)$. $P$ and $1-P$ themselves transform by conjugation under
$SU(3)$. The projector $1-P$ will be important when we discuss Yukawa terms.

%%%%%%%%%%%%%%%%%%%%%%%%%%%%%%%%%%%%%%%%%%%%%%%%%%%%%%%%%%%%%%%%%%%%%%%%%
\section{Squashed 3-Spheres}
%%%%%%%%%%%%%%%%%%%%%%%%%%%%%%%%%%%%%%%%%%%%%%%%%%%%%%%%%%%%%%%%%%%%%%%%%

Let $z = (z_1, z_2)$ be standard coordinates on $\C^2$, and
$ds^2 = d\bz dz$ the Euclidean metric on $\C^2$. Useful polar coordinates are
\be
z_1 = r \cos \half\theta \, e^{i(\chi - \varphi)/2} \,, \quad
z_2 = r \sin \half\theta \, e^{i(\chi + \varphi)/2} \,,
\label{C2polars}
\ee
with ranges $0 \le r$, $0 \le \theta \le \pi$, $0 \le \varphi < 2\pi$ and
$0 \le \chi < 4\pi$. The submanifold of $\C^2$ at fixed $r>0$ is a
round 3-sphere of radius $r$, which in turn is a circle bundle (a Hopf
bundle) over a round 2-sphere, with $z_2/z_1 = \tan \half\theta \,
e^{i\varphi}$ the stereographic coordinate on the 2-sphere, and $\chi$ the
circle coordinate. In polars, the metric $ds^2 = d\bz dz$ becomes
\be
ds^2 = dr^2 + \quart r^2 \left[ d\theta^2 + \sin^2\theta \, d\varphi^2
+ (d\chi - \cos\theta \, d\varphi)^2 \right] ,
\ee
whose angular part is the metric of the round 3-sphere of radius
$r$. Each of these 3-spheres is a $U(2)$ orbit, and there is a unique
$U(2)$ fixed point at $r=0$. Note that the ratio of the circle radius
to the 2-sphere radius is a constant, independent of $r$.

Let us now compare the Fubini--Study metric on $\CP^2$. Using the
inhomogeneous coordinates $Z = (1, z_1, z_2)$, and hence $dZ = (0,
dz_1, dz_2)$, the metric (\ref{FShom}) becomes
\be
ds^2 = \frac{(1+\bz z) d\bz dz
  - (d\bz z)(\bz dz)}{(1 + \bz z)^2} \,,
\label{FSinhom}
\ee
and using the polar coordinates (\ref{C2polars}),
\be
ds^2 = \frac{dr^2}{(1+r^2)^2} + \quart \frac{r^2}{(1+r^2)^2}
\left[ (1+r^2)(d\theta^2 + \sin^2\theta \, d\varphi^2)
  + (d\chi - \cos\theta \, d\varphi)^2 \right] .
\ee
Here, the angular part is the metric of a squashed 3-sphere, a Hopf circle
bundle over a round 2-sphere, where the ratio of the circle radius
to the 2-sphere radius is $1/\sqrt{1+r^2}$, i.e. dependent on $r$.
For us, the important observation is that for small positive $r$, the
radius ratio is close to unity, so the squashed 3-sphere is
approximately round. The consequence for our model is that the custodial
$SO(4)$ symmetry -- the symmetry of a round 3-sphere -- will be
approximately realised.

As $r \to \infty$, the radius ratio approaches zero, compatibly with the
completion of $\CP^2$ at infinity being simply a round 2-sphere, which
is a special 2-dimensional orbit of $U(2)$. This 2-sphere has radius
$\half$ and is at a finite distance from any other point in $\CP^2$.

%%%%%%%%%%%%%%%%%%%%%%%%%%%%%%%%%%%%%%%%%%%%%%%%%%%%%%%%%%%%%%%%%%%%%%%%%
\section{Higgs Kinetic Term}
%%%%%%%%%%%%%%%%%%%%%%%%%%%%%%%%%%%%%%%%%%%%%%%%%%%%%%%%%%%%%%%%%%%%%%%%%

By the Higgs kinetic term we mean the combination of terms involving the
Higgs field's gauge-covariant spacetime derivatives. In our model, the
Higgs field is a map $\Phi$ from spacetime to the target $\CP^2$. Its
derivatives lie in the tangent space to $\CP^2$, and we use the Fubini--Study
metric to construct the Higgs kinetic term, which is quadratic in these
derivatives. Spacetime indices are contracted with the
Minkowski metric $(+,-,-,-)$ using the usual index raising convention.

The Higgs field in homogeneous coordinates is denoted $\Phi =
(\Phi_0, \Phi_1, \Phi_2)$, and in its inhomogeneous form as
\be
\Phi = (M, \phi_1, \phi_2) = (M, \phi) \,,
\ee
where $\phi = (\phi_1, \phi_2)$ is a complex scalar doublet.\footnote{
All these multiplets, and the left-handed fermion doublets below, are column
vectors.} The Higgs field has mass dimension $1$, so the positive
constant $M$ is a mass parameter. We will see later that the expansion
parameter for BSM terms in our model is $\frac{v^2}{M^2}$, where $v$ is
the Higgs vev, so we will need $M \gg v$.

As mentioned earlier, the $SU(3)$ symmetry is explicitly
broken. This is because the point $(M, 0, 0)$ (equivalently $(1,0,0)$)
in $\CP^2$ plays a preferred role. The isotropy group of this point is the
$U(2)$ subgroup with typical element
\be
U = \begin{pmatrix}
    e^{-i\alpha/3} & 0 \\ 0 & e^{i\alpha/6} u
    \end{pmatrix}
\label{SU2U1}
\ee
where $u \in SU(2)$. Its action on the (inhomogeneous) Higgs field is
\be
\Phi \to U\Phi = (M, e^{i\alpha/2}u\phi) \,.
\ee
In our model, as in the SM, it is this $U(2)$ group that is gauged.

We introduce the $SU(2)$ and $U(1)$ gauge potentials, respectively
$A_\mu$ and $B_\mu$, with $A_\mu$ hermitian and $B_\mu$ real. Using
the generators $I_a = \half \sigma_a$ for the Lie algebra of $SU(2)$,
with $\sigma_a$ the Pauli matrices, we have the expansion $A_\mu =
A_\mu^a I_a$ with $A_\mu^a$ real. $I_a$ are the weak isospin operators
and the $U(1)$ generator, the hypercharge operator, is denoted $Y$.
The hypercharge $y$ of a field or particle is its
$Y$-eigenvalue\footnote{We use the `half-scale' convention for
hypercharge, as in  e.g. ref.\cite{PS}.}, and the electric charge
$Q$ is the eigenvalue of $I_3 + Y$. 

In inhomogeneous coordinates, the derivatives and gauge potentials
act non-trivially only on the lower doublet of fields $\phi$, i.e.
\be
D_\mu \Phi = (0, D_\mu \phi)
\ee
where
\be
D_\mu \phi = \pr_\mu \phi - ig A_\mu \phi - \half ig' B_\mu \phi \,,
\ee
and $g$ and $g'$ are the usual gauge coupling constants. The Higgs field
$\phi$ has hypercharge $y=\half$, so its complex components $\phi_1$ and
$\phi_2$ have electric charges $1$ and $0$, respectively.

The Higgs kinetic term is then constructed using the metric
(\ref{FShom}), with $Z$ and $dZ$ substituted by $\Phi$ and $D_\mu\Phi$.
The result is   
\be
\Lag_{\rm Higgs \, Kinetic} = M^2
\frac{(M^2 + \bphi \phi) \overline{D_\mu \phi} D^\mu \phi
  - (\overline{D_\mu \phi} \, \phi)(\bphi D^\mu \phi)}
{\left(M^2 + \bphi \phi \right)^2} \,,
\label{Higgskin}
\ee
where the prefactor $M^2$ is required to obtain an expression of mass
dimension $4$. This kinetic term has novel features compared with
the SM kinetic term $\overline{D_\mu \phi} D^\mu \phi$,
but reduces to it if $|\phi|$ is small compared with $M$.

%%%%%%%%%%%%%%%%%%%%%%%%%%%%%%%%%%%%%%%%%%%%%%%%%%%%%%%%%%%%%%%%%%%%%%%%%
\section{Higgs Potential}
%%%%%%%%%%%%%%%%%%%%%%%%%%%%%%%%%%%%%%%%%%%%%%%%%%%%%%%%%%%%%%%%%%%%%%%%%

The Higgs potential on $\CP^2$ is a real, non-negative function of the
homogeneous Higgs field $\Phi$ that is unchanged by the scaling
$\Phi \to \kappa \Phi$, and is $U(2)$ gauge invariant. It
needs to break the $SU(3)$ symmetry of $\CP^2$, otherwise it would be
constant, and it needs to have its minimum along a non-trivial, squashed
3-sphere orbit of $U(2)$, in order to spontaneously break the
gauge symmetry to $U(1)_{\rm em}$ and for the Higgs mechanism to operate.

An acceptable, simple potential with these properties is
\be
V(\Phi) = \lambda \left(\frac{\bPhi W \Phi}{\bPhi \Phi}\right)^2
\ee
with $W$ a diagonal $3 \times 3$ matrix having its lower pair of entries
equal, and $\lambda$ a dimensionless positive coefficient. Specifically,
evaluating this for the inhomogeneous $\Phi = (M, \phi)$, and with
\be
W = \begin{pmatrix}
    \half v^2 & 0 & 0 \\ 0 & -M^2 & 0 \\ 0 & 0 & -M^2 
    \end{pmatrix} \,,
\ee
we obtain
\be
\Lag_{\rm Higgs \, Potential} = - V(\phi)
= - \lambda M^4 \left( \frac{\half v^2 - \bphi \phi}
{M^2 + \bphi \phi} \right)^2 \,,
\label{Higgspot}
\ee
which has mass dimension $4$. For $|\phi| \ll M$, $V(\phi)$
simplifies to its Standard Model form
$\lambda \left( \half v^2 - \bphi \phi \right)^2$. 

The minimal, vacuum value of $V$ is zero, where $\bphi \phi = \half v^2$. This
is as in the Standard Model, but here the Higgs vacuum manifold is
geometrically a squashed 3-sphere, because of the form of the Higgs kinetic
term (\ref{Higgskin}). $V$ has a local maximum
value $\quart \lambda v^4$ at the point $\phi = 0$, and a further local
maximum value $\lambda M^4$ on the 2-sphere at infinity.

As usual, in unitary gauge, the vacuum field configuration is selected to be
$\phi = (\phi_1, \phi_2) = \left( 0, \frac{1}{\sqrt{2}} v \right)$,
which is charge-neutral and real. The unbroken gauge group is then the
electromagnetic $U(1)_{\rm em}$ subgroup of $U(2)$ with typical element
\be
u_{\rm em} = \begin{pmatrix}
             e^{i\beta} & 0 \\ 0 & 1
             \end{pmatrix} \,.
\ee

%%%%%%%%%%%%%%%%%%%%%%%%%%%%%%%%%%%%%%%%%%%%%%%%%%%%%%%%%%%%%%%%%%%%%%%%%
\section{Gauge Field Tensors}
%%%%%%%%%%%%%%%%%%%%%%%%%%%%%%%%%%%%%%%%%%%%%%%%%%%%%%%%%%%%%%%%%%%%%%%%%

The gauge field tensors and their contribution to the Lagrangian are
exactly as in the Standard Model. The $SU(2)$ and $U(1)$
field tensors are
\bea
F_{\mu\nu} &=& \pr_\mu A_\nu - \pr_\nu A_\mu - ig[A_\mu, A_\nu] \,, \\
B_{\mu\nu} &=& \pr_\mu B_\nu - \pr_\nu B_\mu \,.
\eea 
Using the expansion of the $SU(2)$ gauge potential,
$A_\mu = A_\mu^a I_a$, the $SU(2)$ field tensor acquires the
expansion $F_{\mu\nu} = F_{\mu\nu}^a I_a$, where 
\be
F_{\mu\nu}^a = \pr_\mu A_\nu^a - \pr_\nu A_\mu^a
+ g\e^{abc} A_\mu^b A_\nu^c \,.
\ee
The field tensor terms in the Lagrangian are
\be
\Lag_{\rm Field \, Tensor} = -\quart F_{\mu\nu}^a F^{\mu\nu a} - \quart
B_{\mu\nu} B^{\mu\nu} \,.
\ee

%%%%%%%%%%%%%%%%%%%%%%%%%%%%%%%%%%%%%%%%%%%%%%%%%%%%%%%%%%%%%%%%%%%%%%%%%
\section{Fermion Kinetic Terms}
%%%%%%%%%%%%%%%%%%%%%%%%%%%%%%%%%%%%%%%%%%%%%%%%%%%%%%%%%%%%%%%%%%%%%%%%%

We start from the trivial $\C^3$ bundle over $\CP^2$, whose constant fibre
$\V = \C^3$ is unaffected by changes in the value of the Higgs field
$\Phi$, i.e. the point in $\CP^2$. We next exploit the explicit breaking
of $SU(3)$ by the choice of vector $(1,0,0) \in \C^3$, which restricts
the gauge group to $U(2)$ with group elements (\ref{SU2U1}).
This allows us to pick out the 2-dimensional subspace $\V_L$ of vectors
in $\C^3$ whose upper component is zero, and the 1-dimensional subspace
$\V_R$ of vectors whose lower two components are zero. Finally, using this
gauge-invariant splitting, we assume that the left-handed fermion fields
are (electroweak) doublets lying in $\V_L$ and the right-handed fermion
fields are singlets lying in $\V_R$. $SU(2)$ acts canonically
from the left on $\V_L$ and trivially on $\V_R$. The action of the $U(1)$
generator $Y$ has more freedom compatible with gauge invariance. Its
eigenvalues -- the hypercharges of the fermion fields -- do not have
to be those of the generating matrix for the group elements
(\ref{SU2U1}), namely $-\frac{1}{3}$ and $\frac{1}{6}$.

In the context of the Standard Model, the above remarks are
a restatement of what is well known. The reason for making
them is that there is an alternative possibility, exploiting the
Higgs field $\Phi$. We could have split the trivial $\C^3$ bundle
over $\CP^2$ into the 1-dimensional tautological bundle and the
2-dimensional tangent bundle, achieved at each point of $\CP^2$
by using the projection operators $P$ and
$1-P$, as described in section 2. We could then have
assumed that the left-handed fermion fields lie in the tangent bundle,
and the right-handed fields in the tautological bundle. This
assumption would have been compatible with retaining the full $SU(3)$
symmetry. However, it would have made the fermions dependent on the
Higgs field at the kinematic level. We find this unattractive. It also
makes construction of Yukawa interaction terms problematic. Instead, we
have exploited the constant vector $(1,0,0)$ to produce the
doublet/singlet fermion field structure. The projection operator
$1-P$ will later be key to the construction of gauge-invariant Yukawa terms.

These assumptions mean that the kinetic terms (involving gauge-covariant
spacetime derivatives) for the fermions are exactly as in the Standard
Model, making use of the Dirac matrices $\g^\mu$ and $\g^5$. Fermion
fields are doublet 4-spinors $\psi_L$ and singlet 4-spinors $\psi_R$
satisfying, respectively, $\g^5 \psi_L = -\psi_L$ and
$\g^5 \psi_R = \psi_R$.

For the lightest, first-generation quark pair, the u-quark and d-quark,
the fields are the left-handed doublet $q_L = (u_L, d_L)$ and the two
right-handed singlets $u_R$ and $d_R$. The second- and third-generation
quark pairs are similar and we do not write their Lagrangians
explicitly. The quark kinetic terms are
\be
\Lag_{\rm Quark \, Kinetic} = \overline{q_L} \, i\g^\mu D_\mu q_L
+ \overline{u_R} \, i\g^\mu D_\mu u_R
+ \overline{d_R} \, i\g^\mu D_\mu d_R \,,
\label{QuarkKin}
\ee
where the different gauge-covariant derivatives are
\bea
D_\mu q_L &=& \pr_\mu q_L - ig A_\mu q_L - ig' y_L^q B_\mu q_L \,, \\
D_\mu u_R &=& \pr_\mu u_R - ig' y_R^u B_\mu u_R \,, \\
D_\mu d_R &=& \pr_\mu d_R - ig' y_R^d B_\mu d_R
\eea
and the quark hypercharges are $(y_L^q, y_R^u, y_R^d) =
(\frac{1}{6}, \frac{2}{3}, -\frac{1}{3})$. Note that the overbar on
the fermion fields in (\ref{QuarkKin}) is the Dirac conjugate,
including $\g^0$.

For the first-generation lepton pair, the electron and its neutrino,
the fields are the doublet $l_L = (\nu_L, e_L)$ and the one singlet $e_R$.
For the similar second- and third-generation lepton pairs we again do not
write the Lagrangians explicitly. The lepton kinetic terms are
\be
\Lag_{\rm Lepton \, Kinetic} = \overline{l_L} \, i\g^\mu D_\mu l_L
+ \overline{e_R} \, i\g^\mu D_\mu e_R \,,
\ee
with the gauge-covariant derivatives
\bea
D_\mu l_L &=& \pr_\mu l_L - ig A_\mu l_L - ig' y_L^l B_\mu l_L \,, \\
D_\mu e_R &=& \pr_\mu e_R - ig' y_R^e B_\mu e_R  
\eea
and the lepton hypercharges $(y_L^l, y_R^e) = (-\half, -1)$.

These fermions have electric charges $(Q^u, Q^d, Q^\nu, Q^e) =
(\frac{2}{3}. -\frac{1}{3}, 0, -1)$. 

%%%%%%%%%%%%%%%%%%%%%%%%%%%%%%%%%%%%%%%%%%%%%%%%%%%%%%%%%%%%%%%%%%%%%%%%%
\section{Yukawa Terms}
%%%%%%%%%%%%%%%%%%%%%%%%%%%%%%%%%%%%%%%%%%%%%%%%%%%%%%%%%%%%%%%%%%%%%%%%%

The Yukawa terms in the Standard Model generate
gauge-invariant masses for the fermions. They couple the Higgs field to
fermion bilinears, and the masses generated are proportional to a
Yukawa coupling times the Higgs vev $v$. We first recall the two ways for
contracting the left-handed fermion doublets with the Higgs doublet
field $\phi$, in order to have $SU(2)$ singlets. Let us illustrate
this for the left-handed quark doublet $q_L$. In components,
\be
q_L = (u_L, d_L) \quad {\rm and} \quad \phi = (\phi_1, \phi_2) \,.
\ee
The first contraction is the obvious $\overline{q_L} \, \phi$,
and the Yukawa term obtained from this is
\be
-\lambda_d \, \overline{q_L} \, \phi \, d_R
= -\lambda_d (\overline{u_L} \, \phi_1 \, d_R
+ \overline{d_L} \, \phi_2 \, d_R) \,,
\ee
which is a $U(2)$ singlet because of the quark and
Higgs field hypercharge assignments. Only a d-quark mass is generated
through this term in the Higgs vacuum, $(\phi_1, \phi_2) =
\left( 0, \frac{1}{\sqrt{2}} v \right)$. To generate a u-quark mass,
one needs to exploit the conjugate of the Higgs doublet,
$\phi_C = i\sigma_2 \phi^* = (\phi_2^*, -\phi_1^*)$. This is also a
doublet of $SU(2)$, like $\phi$, but has the opposite hypercharge. The
second contraction with $q_L$ is then $\overline{q_L} \, \phi_C$, and
from this one obtains the Yukawa term
\be
-\lambda_u \, \overline{q_L} \, \phi_C \, u_R
= -\lambda_u (\overline{u_L} \, \phi_2^* \, u_R
- \overline{d_L} \, \phi_1^* \, u_R) \,.
\ee
$\lambda_d$ and $\lambda_u$ are the Yukawa coupling constants that
combine with the Higgs vev to give the d- and u-quark masses
$\frac{1}{\sqrt{2}} \lambda_d v$ and $\frac{1}{\sqrt{2}} \lambda_u v$.

For the leptons there is just one Yukawa term, giving the electron a
mass. Since $e_L$ is the lower component of its doublet, the required term
is
\be
-\lambda_e \, \overline{l_L} \, \phi \, e_R
= -\lambda_e (\overline{\nu_L} \, \phi_1 \, e_R
+ \overline{e_L} \, \phi_2 \, e_R) \,,
\ee
the analogue of the term giving the d-quark mass. The electron mass is
$\frac{1}{\sqrt{2}} \lambda_e v$. The neutrino is left-handed
and massless, with no right-handed field, so there is no need to
use $\phi_C$ again. 

It is rather mysterious that the Higgs field occurs in two
different ways, but it is straightforward to verify directly that
$\phi$ and $\phi_C$ transform the same way under the left-action
of $SU(2)$. This can also be understood by considering left
multiplication by the $SU(2)$ matrix
\be
u = \begin{pmatrix}
  a & b \\
  -b^* & a^*
\end{pmatrix}
\ee
(with $|a|^2 + |b|^2 = 1$) on the $2 \times 2$ Higgs field matrix
\be
\phi_m = \begin{pmatrix}
         \phi_2^* & \phi_1 \\
         -\phi_1^* & \phi_2
       \end{pmatrix} \,,
\ee
with columns $\phi_C$ and $\phi$. $\phi_m$ is an $SU(2)$ matrix times
the scalar multiple $\sqrt{|\phi_1|^2 + |\phi_2|^2}$. The product
$\phi'_m = u\phi_m$ is consequently another $SU(2)$ matrix times the
same multiple and has the same form, with $(\phi_1, \phi_2)$ replaced by
\be
(\phi'_1, \phi'_2) = (a \phi_1 + b \phi_2,
-b^* \phi_1 + a^* \phi_2) \,.
\ee
Matrix multiplication by $u$ acts the same way on the left and right
columns of $\phi_m$, and therefore $\phi$ and $\phi_C$ transform the
same way under $SU(2)$. 

We now turn to the $\CP^2$ model. There is still no unification of
$\phi$ and $\phi_C$, and two distinct types of Yukawa term are needed.
We will use the homogeneous Higgs field $\Phi$ in our Yukawa
terms, and must ensure that the result is gauge invariant and
unchanged by rescaling by $\kappa$. The useful
scale-invariant objects derived from $\Phi$ are the projectors
\be
P = \frac{\Phi\bPhi}{\bPhi\Phi} \quad {\rm and} \quad
1-P = \frac{\bPhi\Phi-\Phi\bPhi}{\bPhi\Phi} \,.
\ee
In terms of the components of the inhomogeneous Higgs field
$\Phi = (M, \phi_1, \phi_2)$, these projectors are the
$3 \times 3$ matrices
\be
P = \frac{1}{M^2 + |\phi_1|^2 +|\phi_2|^2}
  \begin{pmatrix}
  M^2 & M\phi_1^* & M\phi_2^* \\
  M\phi_1 & |\phi_1|^2 & \phi_1\phi_2^* \\
  M\phi_2 & \phi_2\phi_1^* & |\phi_2|^2
  \end{pmatrix}
\ee
and
\be
1-P = \frac{1}{M^2 + |\phi_1|^2 + |\phi_2|^2}
  \begin{pmatrix}
  |\phi_1|^2 + |\phi_2|^2 & -M\phi_1^* & -M\phi_2^* \\
  -M\phi_1 & M^2 + |\phi_2|^2 & -\phi_1 \phi_2^* \\
  -M\phi_2 & -\phi_2 \phi_1^* & M^2 + |\phi_1|^2
\end{pmatrix} .
\ee

Now recall that in our model the left-handed quark doublet and
the right-handed d-quark singlet are (column) 3-vector fields
$q_L = (0,u_L,d_L)$ and $(d_R,0,0)$. Acting with $1-P$ on the
d-quark singlet gives the 3-vector
\be
\frac{1}{M^2 + |\phi_1|^2 + |\phi_2|^2}
\biggl( (|\phi_1|^2 + |\phi_2|^2)d_R, \, -M\phi_1 \, d_R,
\, -M\phi_2 \, d_R \biggr) .
\ee
Contracting with $q_L$, and including an appropriate coupling
constant, gives the Yukawa term
\be
\frac{-\Lambda_d M^2}{M^2 + |\phi_1|^2 + |\phi_2|^2}
(\overline{u_L} \, \phi_1 \, d_R + \overline{d_L} \, \phi_2 \, d_R) \,,
\ee
where we have included an extra factor of $M$ so that this term has mass
dimension 4. The Yukawa coupling $\Lambda_d$ is dimensionless and
close to its value in the Standard Model. When the Higgs
field acquires its vev, the d-quark mass becomes
\be
m_d = \frac{1}{\sqrt{2}} \frac{M^2}{M^2 + \half v^2} \Lambda_d v \,.
\ee
The Yukawa term for the lepton doublet is similar, giving an electron
mass
\be
m_e = \frac{1}{\sqrt{2}} \frac{M^2}{M^2 + \half v^2} \Lambda_e v \,.
\ee

Note that the projector $1-P$ acting on a right-handed fermion singlet
produces a doublet and a singlet. The doublet plays the role in the
$\CP^2$ model that the product of $\phi$ with the right-handed singlet
does in the Standard Model, and its contraction with the corresponding
left-handed fermion doublet is $U(2)$ invariant. This is the term we
use. One could generate further Yukawa terms by using the projected
singlet, but these would be of higher order in $\phi$.

For the second type of Yukawa term, giving the u-quark its mass, we need to
replace $\phi$ by its conjugate $\phi_C$, and use this to construct
the conjugate projector $1 - P_C$. The right-handed u-quark singlet
is the 3-vector $(u_R, 0, 0)$. Acting with $1-P_C$ on this, and
again contracting with $q_L$, gives the Yukawa term
\be
\frac{-\Lambda_u M^2}{M^2 + |\phi_1|^2 + |\phi_2|^2}
(\overline{u_L} \, \phi_2^* \, u_R - \overline{d_L} \, \phi_1^* \, u_R) \,.
\ee
The u-quark mass is then
\be
m_u = \frac{1}{\sqrt{2}} \frac{M^2}{M^2 + \half v^2} \Lambda_u v \,.
\ee

The total contribution of the Yukawa terms to the Lagrangian of the
$\CP^2$ model is
\bea
\Lag_{\rm Yukawa} &=& - \, \frac{M^2}{M^2 + |\phi_1|^2 + |\phi_2|^2}
\Bigl[ \Lambda_d (\overline{u_L} \, \phi_1 \, d_R + \overline{d_L} \,
\phi_2 \, d_R) \nn \\
&& \quad
+ \Lambda_u (\overline{u_L} \, \phi_2^* \, u_R - \overline{d_L} \,
\phi_1^* \, u_R) 
+ \Lambda_e (\overline{\nu_L} \, \phi_1 \, e_R + \overline{e_L} \,
\phi_2 \, e_R) \Bigr] .
\eea

%%%%%%%%%%%%%%%%%%%%%%%%%%%%%%%%%%%%%%%%%%%%%%%%%%%%%%%%%%%%%%%%%%%%%%%%%
\section{Phenomenology Beyond the Standard Model}
%%%%%%%%%%%%%%%%%%%%%%%%%%%%%%%%%%%%%%%%%%%%%%%%%%%%%%%%%%%%%%%%%%%%%%%%% 

The Lagrangian of our proposed $\CP^2$ SMEFT is
\bea
\Lag &=& \Lag_{\rm Field \, Tensor} + \Lag_{\rm Higgs \, Kinetic}
+ \Lag_{\rm Higgs \, Potential} \nn \\
&& \qquad + \, \Lag_{\rm Quark \, Kinetic} 
+ \Lag_{\rm Lepton \, Kinetic} + \Lag_{\rm Yukawa} \,,
\label{LagSMEFT}
\eea
with its contributions defined in sections 5 -- 8 above. Some of the apparent
differences from the Standard Model can be removed by a recalibration of
parameters and rescaling of the fields. But there remain some
significant modifications to SM relations, and we will use these
together with recent experimental results as reviewed by the Particle
Data Group (PDG) \cite{PDG} to place
constraints on the BSM mass parameter $M$. Also, when the denominator
factors are expanded out, there arise several dimension-6 terms
that are SM terms multiplied by $\bphi \phi$. In the
following, we will explore the consequences of the $\CP^2$ SMEFT for
(i) gauge boson masses, (ii) Higgs cubic and quartic couplings, (iii)
Yukawa couplings, and (iv) Higgs boson couplings to gauge bosons.
Constraints on $M$ currently come mainly from
the $W$-boson to $Z$-boson mass ratio.

\subsection{The $W$-boson to $Z$-boson mass ratio}

The gauge boson masses are determined using the terms in $\Lag$
involving the gauge field tensor and covariant derivatives of the
Higgs field. Most interesting is the novel contribution
$(\overline{D_\mu \phi} \, \phi)(\bphi D^\mu \phi)$ to the Higgs
kinetic term. 

The gauge couplings $g$ and $g'$ are the same as in the Standard
Model, as they control the tree-level couplings of gauge
bosons to fermions. The weak mixing angle $\theta_w$ is therefore
also the same; it is defined, as usual, via
\be
\cos \theta_w = \frac{g}{\sqrt{g^2 + g'^2}} \,, \quad
\sin \theta_w = \frac{g'}{\sqrt{g^2 + g'^2}} \,.
\ee
The field tensor term in the Lagrangian (\ref{LagSMEFT}) has its
canonical form, so the
(squared) spacetime derivatives of the gauge boson fields have
canonical normalisation. The gauge boson masses are then extracted
from the gauge-covariant Higgs kinetic term (\ref{Higgskin}) by
setting the Higgs field equal to its vacuum value $\phi = \left( 0,
  \frac{1}{\sqrt{2}} v \right)$, and ignoring the derivatives of
$\phi$. The two contributions (without their prefactors) are
\bea
\overline{D_\mu \phi} D^\mu \phi &=& \frac{v^2}{8}
\left[ g^2(A_\mu^1)^2 + g^2(A_\mu^2)^2 + (gA_\mu^3 - g'B_\mu)^2
\right] \,, \nn \\ 
(\overline{D_\mu \phi} \, \phi)(\bphi D^\mu \phi) &=&
\frac{v^4}{16}(gA_\mu^3 - g'B_\mu)^2 \,,
\label{gaugebosonmass}
\eea
(where $(A_\mu^1)^2$ is shorthand for $A_\mu^1 A^{1 \mu}$,
etc.). The second of these does not appear at all in the Standard Model
Lagrangian. It is the term denoted ${\cal Q}_{\phi D}$ \cite{Grz} or
${\cal O}_{\phi D}$ \cite{Blas} in the classification of dimension-6
SMEFT operators, and denoted ${\cal O}_{HD}$ \cite{Ellis1, Ellis2}
in the analysis of constraints on their coefficients.

As usual, we define the $Z$-boson field and photon field to be
\be
Z_\mu = \frac{1}{\sqrt{g^2 + g'^2}}(gA_\mu^3 - g'B_\mu) \,,
\quad A_\mu = \frac{1}{\sqrt{g^2 + g'^2}}(g'A_\mu^3 + gB_\mu) \,,
\ee
or equivalently
\be
Z_\mu = \cos \theta_w \, A_\mu^3 - \sin \theta_w \, B_\mu \,,
\quad A_\mu = \sin \theta_w \, A_\mu^3 + \cos \theta_w \, B_\mu \,.
\ee
The $W$-boson fields are $W_\mu^\pm = \frac{1}{\sqrt{2}}
(A_\mu^1 \mp i A_\mu^2)$. It is not surprising that both gauge-boson mass
terms (\ref{gaugebosonmass}) have no contribution from the photon field.
The photon remains massless because of the unbroken $U(1)_{\rm em}$ symmetry.

The combined gauge-boson mass term, including the prefactors, is
\bea
&& \frac{M^2}{\left(M^2 + \half v^2 \right)^2}
\biggl[ \left( M^2 + \half v^2 \right)\frac{v^2}{8}
\left[ 2g^2 W_\mu^- W^{\mu +} + (g^2 + g'^2)Z_\mu Z^\mu \right] \nn \\
&& \qquad\qquad\quad\qquad\qquad\qquad
-\frac{v^4}{16}(g^2 + g'^2)Z_\mu Z^\mu \biggr] .
\eea
$v^4 Z_\mu Z^\mu$ cancels here, and what remains can be reexpressed as
\be
\frac{M^2}{\left(M^2 + \half v^2 \right)^2}
  \left(\frac{gv}{2} \right)^2 
  \left[ \left( M^2 + \half v^2 \right) W_\mu^- W^{\mu +}
  + \half \frac{M^2}{\cos^2 \theta_w} Z_\mu Z^\mu \right] .
\ee
From this we read off the gauge boson masses
\be
m_W = \frac{M}{\sqrt{M^2 + \half v^2}}
\left(\frac{gv}{2} \right) \,, \quad
m_Z = \frac{M^2}{\left(M^2 + \half v^2 \right)}
\left(\frac{gv}{2} \right) \frac{1}{\cos \theta_w} \,.
\ee

From here on, we work to leading non-trivial order in $\frac{v^2}{M^2}$,
as the estimates below show that this quantity is less than $0.01$. The
masses become
\be
m_W = \left( 1 - \frac{v^2}{4M^2} \right)\left(\frac{gv}{2} \right) \,,
\quad m_Z = \left( 1 - \frac{v^2}{2M^2} \right)
\left(\frac{gv}{2} \right) \frac{1}{\cos \theta_w} \,.
\label{gaugebosonmasses}
\ee
The Higgs vev $v$ can be expressed in terms of the conventional
$v_{\rm SM} = 246.22$ GeV to match the $Z$-boson mass formula in
the Standard Model. To do this, we define
\be
v_{\rm SM} = \left( 1 - \frac{v^2}{2M^2} \right) v
\label{vev}
\ee
and find
\be
m_W = \left( 1 + \frac{v_{\rm SM}^2}{4M^2} \right)
\left(\frac{gv_{\rm SM}}{2} \right) \,,
\quad m_Z = \left(\frac{gv_{\rm SM}}{2} \right) \frac{1}{\cos \theta_w} \,.
\ee
The key result here is that the squared $W$-boson to $Z$-boson mass
relation is
\be
m_W^2 = \left( 1 + \frac{v_{\rm SM}^2}{2M^2} \right) m_Z^2
\cos^2 \theta_w \,.
\ee
This is a small modification of the Standard Model tree level relation
$m_W^2 = m_Z^2 \cos^2 \theta_w$ in the direction of increasing
the $W$-boson mass towards the $Z$-boson mass.

There are SM loop corrections to this tree level relation, but we can
assume that these are essentially the same in the $\CP^2$ SMEFT. The
$\CP^2$ SMEFT therefore predicts a BSM ``new physics'' factor
\be
\rho_0 = 1 + \frac{v_{\rm SM}^2}{2M^2} \,.
\label{rho0}
\ee
Current data \cite{PDG} gives $\rho_0 = 1.00031 \pm 0.00019$ if
the combined CERN and Fermilab D0 result of 80.360 GeV for the
$W$-boson mass is used, i.e. slightly higher than the SM prediction of
$\rho_0 = 1$ but of little statistical significance.
Any BSM correction to the SM is therefore small, but appears to
be in the direction of an increase of the $W$-boson mass relative to
the $Z$-boson mass, as we find in the $\CP^2$ SMEFT. At a confidence
level of $2\sigma$, $\rho_0 < 1.00031 + 0.00038 = 1.00069$ and
formula (\ref{rho0}) implies that with the same confidence, the
$\CP^2$ SMEFT requires $M > 27 \, v_{\rm SM} \simeq 6.6$ TeV.

The Fermilab CDF result for the $W$-boson mass, based on a recent
reanalysis of older data \cite{CDF}, is 80.434 GeV, with an
error estimate making this value $5\sigma$ higher than the combined CERN/D0
result and $7\sigma$ higher than the SM prediction. If we use this,
then $\rho_0 \simeq 1.002$, and from (\ref{rho0})
we deduce that $M \simeq 4$ TeV. Within our model, this estimate for
$M$ is not ruled out by the Yukawa couplings, to be discussed below.

%%%%%%%%%%%%%%%%%%%%%%%%%%%%%%%%%%%%%%%%%%%%%%%%%%%%%%%%%%%%%%%%%%%%%%%%% 
\subsection{Higgs-boson mass and couplings}
%%%%%%%%%%%%%%%%%%%%%%%%%%%%%%%%%%%%%%%%%%%%%%%%%%%%%%%%%%%%%%%%%%%%%%%%%

Here we consider the terms of the Lagrangian involving the Higgs field
alone, leaving out couplings of the Higgs field to the gauge bosons
and fermions.

In the SM, these terms are
\be
\Lag_{\rm Higgs} = \overline{\pr_\mu \phi} \pr^\mu \phi
- \lambda \left( \half v_{\rm SM}^2 - \overline{\phi} \phi \right)^2 \,,
\ee
and in unitary gauge
\be
\phi = \frac{1}{\sqrt{2}} (0, v_{\rm SM}+H) \,,
\ee
where $v_{\rm SM}$ is the SM Higgs vev and $H$ the physical Higgs boson
field. Expanding out, we obtain
\be
\Lag_{\rm Higgs} = \half \pr_\mu H \pr^\mu H
- \lambda v_{\rm SM}^2 H^2 -
\lambda v_{\rm SM} H^3 - \quart \lambda H^4 \,.
\ee
A standard notation for this is
\be
\Lag_{\rm Higgs} = \half \pr_\mu H \pr^\mu H - \half m_H^2 H^2 -
\frac{1}{6} \lambda_3 H^3 - \frac{1}{24} \lambda_4 H^4 \,,
\label{CanonHiggs}
\ee
so $m_H^2 = 2\lambda v_{\rm SM}^2$ is the Higgs boson mass,
$\lambda_3 = 6\lambda v_{\rm SM} = 3\frac{m_H^2}{v_{\rm SM}}$ is the
cubic Higgs coupling, and $\lambda_4 = 6\lambda = 3\frac{m_H^2}{v_{\rm SM}^2}$
is the quartic Higgs coupling. As $m_H$ and $v_{\rm SM}$ are precisely
known, the cubic and quartic Higgs couplings have precisely predicted
SM values.

In the $\CP^2$ SMEFT, terms in $\Lag_{\rm Higgs}$ acquire an additional
denominator factor $(1 + \frac{\overline{\phi} \phi}{M^2})^2$. Working
to leading order, i.e. up to dimension 6 terms, we can replace
this by a numerator factor $(1 - 2\frac{\overline{\phi} \phi}{M^2})$.
Then, going to unitary gauge and writing
\be
\phi = \frac{1}{\sqrt{2}} (0, v+h) \,,
\ee
we find from the kinetic and potential contributions (\ref{Higgskin})
and (\ref{Higgspot}) that
\bea
\Lag_{\rm Higgs} &=& \half \left(1 - \frac{v^2}{M^2} -
\frac{2v}{M^2} h - \frac{1}{M^2} h^2 \right) \pr_\mu h \pr^\mu h \nonumber \\
&& \ - \quart \lambda 
\left(1 - \frac{v^2}{M^2} - \frac{2v}{M^2} h - \frac{1}{M^2} h^2
\right) (4v^2 h^2 + 4v h^3 + h^4).
\eea
Note that this includes terms where $\half \pr_\mu h \pr^\mu h$ is
multiplied by $h$ and $h^2$. These are small, momentum dependent corrections
to the cubic and quartic Higgs couplings. From now on, we drop these,
leaving, after collecting terms,
\bea
\Lag_{\rm Higgs} &=& \half \left(1 - \frac{v^2}{M^2} \right) \pr_\mu h
\pr^\mu h - \lambda v^2 \left(1 - \frac{v^2}{M^2} \right) h^2 \nonumber \\
&& \quad - \lambda v \left(1 - \frac{3v^2}{M^2} \right) h^3
- \quart \lambda \left(1 - \frac{13v^2}{M^2} \right) h^4 + O(h^5) \,.
\eea
We need to rescale $h$ to reproduce the canonical kinetic term for the
Higgs boson field. So we define
\be
H = \left(1 - \frac{v^2}{2M^2} \right) h
\ee
and obtain (to leading order in $\frac{v^2}{M^2}$)
\bea
\Lag_{\rm Higgs} &=& \half \pr_\mu H \pr^\mu H - \lambda v^2
H^2 \nonumber \\
&& \quad - \lambda v \left(1 - \frac{3v^2}{2M^2} \right) H^3
- \quart \lambda \left(1 - \frac{11v^2}{M^2} \right) H^4 \,.
\eea
$m_H^2 = 2\lambda v^2$ is therefore the Higgs boson mass, and we can eliminate
$\lambda$ in favour of $m_H$ in the cubic and quartic terms to obtain
\bea
\Lag_{\rm Higgs} &=& \half \pr_\mu H \pr^\mu H - \lambda v^2
H^2 \nonumber \\
&& \quad - \frac{m_H^2}{2v} \left(1 - \frac{3v^2}{2M^2} \right) H^3
- \frac{m_H^2}{8v^2} \left(1 - \frac{11v^2}{M^2} \right) H^4 \,.
\eea
Writing this again in the form (\ref{CanonHiggs}), we read off that
\be
\lambda_3 = \frac{3m_H^2}{v} \left(1 - \frac{3v^2}{2M^2} \right)
\quad {\rm and} \quad
\lambda_4 = \frac{3m_H^2}{v^2} \left(1 - \frac{11v^2}{M^2} \right) \,.
\ee
To compare with the SM couplings we need to make the transformation
(\ref{vev}) from $v$ to $v_{\rm SM}$. To leading order,
\be
\lambda_3 = \frac{3m_H^2}{v_{\rm SM}}
\left(1 - \frac{2v_{\rm SM}^2}{M^2} \right)  \quad {\rm and} \quad
\lambda_4 = \frac{3m_H^2}{v_{\rm SM}^2} \left(1 - \frac{12v_{\rm SM}^2}{M^2}
\right) \,.
\ee
Then, writing these couplings as \cite{LLRZ}
\be
\lambda_3 = \frac{3m_H^2}{v_{\rm SM}} (1 + \kappa_3)
\quad {\rm and} \quad
\lambda_4 = \frac{3m_H^2}{v_{\rm SM}^2} (1 + \kappa_4)
\ee
we read off the BSM modifications
\be
\kappa_3 = -\frac{2v_{\rm SM}^2}{M^2}
\quad {\rm and} \quad  
\kappa_4 = -\frac{12 v_{\rm SM}^2}{M^2} \,.
\ee
The $\CP^2$ SMEFT therefore predicts that both these parameters are
negative and in the ratio $\kappa_4 = 6\kappa_3$.

Unfortunately, the cubic and quartic Higgs couplings are not yet
determined experimentally. There are only weak upper bounds on
their strengths. In future, using the HL-LHC at CERN, it is hoped that
these couplings can be measured and compared with SM predictions.
Even then, the small modifications $\kappa_3$ and $\kappa_4$ to the
SM predictions will probably remain out of reach. So these couplings
give no useful constraint on $M$. However, the discussion here has
clarified the need for the rescaling of the Higgs boson field $h$ as
well as the Higgs vev $v$. These rescalings have an important effect
on the Yukawa couplings, which we calculate next, and on the couplings
of the Higgs boson to the gauge bosons.

%%%%%%%%%%%%%%%%%%%%%%%%%%%%%%%%%%%%%%%%%%%%%%%%%%%%%%%%%%%%%%%%%%%%%%%%% 
\subsection{Higgs-fermion couplings}
%%%%%%%%%%%%%%%%%%%%%%%%%%%%%%%%%%%%%%%%%%%%%%%%%%%%%%%%%%%%%%%%%%%%%%%%%

The $\CP^2$ SMEFT modifies the SM relationship between fermion
masses and the coupling of fermions to the Higgs boson. As we will
see, this modification is by a universal factor for all fermion
types. The coupling controls both the production and decay strengths
for the Higgs boson, and has been accurately measured for the three
fermions of the third generation -- the t-quark and b-quark, and the
$\tau$-lepton. Currently, the SM relationship is confirmed, within
experimental uncertainties, so this places limits on BSM
modifications, and hence a lower bound on $M$ in the $\CP^2$ SMEFT.

The kinetic terms for fermions have their canonical form in $\Lag$, so
we need only investigate the Yukawa terms. To determine the fermion
masses and couplings we need to retain only the terms up to linear order
in the Higgs field $h$. As before, we convert the denominator
factor in $\Lag_{\rm Yukawa}$ to a numerator factor and obtain for
any fermion $f$,
\bea
\Lag_{\rm Yukawa} &=& - \frac{1}{\sqrt{2}} \Lambda_f \overline{f} f
\left( 1 - \frac{v^2}{2M^2} - \frac{v}{M^2} h - \frac{1}{2M^2} h^2
\right) (v+h) \nonumber \\
&=& - \frac{1}{\sqrt{2}} \Lambda_f \overline{f} f
\left(\left(1 - \frac{v^2}{2M^2} \right) v
+ \left(1 - \frac{3v^2}{2M^2} \right) h + O(h^2) \right).
\eea
We next rescale $v$ and $h$ by the common factor
$1- \frac{v^2}{2M^2}$ to obtain the canonical
$v_{\rm SM}$ and $H$. To linear order in $H$, the Yukawa term becomes
\be
\Lag_{\rm Yukawa} = - \frac{1}{\sqrt{2}} \Lambda_f \overline{f} f
\left(v_{\rm SM} + \left( 1 - \frac{v_{\rm SM}^2}{M^2} \right) H
\right).
\ee
The mass of the fermion $f$ and its Yukawa coupling to $H$ are
\be
m_f = \frac{1}{\sqrt{2}} \Lambda_f v_{\rm SM}
\quad {\rm and} \quad
y_f = \frac{1}{\sqrt{2}} \Lambda_f \left(1 - \frac{v_{\rm SM}^2}{M^2}
\right) \,,
\ee
giving the universal BSM ratio
\be
\kappa_f \equiv \frac{y_f}{m_f}  v_{\rm SM} =
1 - \frac{v_{\rm SM}^2}{M^2} \,,
\ee
a slight decrease compared to the SM value $1$.

The measured values of this ratio \cite{PDG} are
$\kappa_t = 1.0 \pm 0.1$, $\kappa_b = 1.0 \pm 0.15$,
$\kappa_\tau = 0.9 \pm 0.1$, which are all consistent with the SM. For
the sake of argument, suppose that more precise measurements give a
universal value $\kappa_f = 0.95$. Then $M =$ O(1 TeV). More realistically,
the current data imply that $M$ is O(1 TeV) or larger in the $\CP^2$
SMEFT. This lower bound is weaker than that obtained from the
$W$-boson to $Z$-boson mass ratio, so it is the mass ratio that
currently is most restrictive. Note that this lower bound does not
conflict with the CDF result for the $W$-boson to $Z$-boson mass
ratio.

\subsection{The $WWH$ and $ZZH$ couplings}

Finally, we determine the tree-level couplings of the $Z$- and $W$-bosons
to the physical Higgs boson. These couplings are found by
expanding the Higgs kinetic Lagrangian (\ref{Higgskin}) to linear
order in the Higgs field $h$. The coupling of the photon to the Higgs boson
vanishes at tree level. As before, we go to unitary gauge,
writing $\phi = \frac{1}{\sqrt{2}}(0, v+h)$, and will rescale both $v$
and the field $h$ to obtain the canonical SM mass for the $Z$-boson and
the canonical kinetic term for the Higgs field. The rescaled quantities are
\be
v_{\rm SM} = \left( 1 - \frac{v^2}{2M^2} \right) v \,, \quad
H = \left(1 - \frac{v^2}{2M^2} \right) h \,.
\label{rescaledvh}
\ee

The desired coupling terms are quadratic in the gauge boson fields and
linear in the Higgs field, and arise from (\ref{Higgskin}) after
excluding the derivative $\pr_\mu h$. We set
\be
M^2 + \bphi \phi = M^2 + \half (v + h)^2
\ee
and from (\ref{Higgskin}) read off the remaining part
\bea
&& \frac{M^2}{\left(M^2 + \half (v+h)^2 \right)^2} \times \nn \\
&& \qquad \biggl[ \left( M^2 + \half (v+h)^2 \right)\frac{(v+h)^2}{8}
\left[ 2g^2 W_\mu^- W^{\mu +} + (g^2 + g'^2)Z_\mu Z^\mu \right] \nn \\
&& \qquad\qquad\qquad
-\frac{(v+h)^4}{16}(g^2 + g'^2)Z_\mu Z^\mu \biggr] ,
\eea
where, similarly as before, the $(v+h)^4 Z_\mu Z^\mu$ contributions
cancel, leaving
\bea
&& \frac{M^2}{\left(M^2 + \half (v+h)^2 \right)^2}
\biggl[ M^2 \frac{(v+h)^2}{8}
\left[ 2g^2 W_\mu^- W^{\mu +} + (g^2 + g'^2)Z_\mu Z^\mu \right] \nn \\
&& \qquad\qquad\quad\qquad\qquad\qquad
+\frac{(v+h)^4}{8} g^2 W_\mu^- W^{\mu +} \biggr] .
\eea
Expanding in powers of $h$, we reproduce at zeroth order the masses of the
gauge bosons, and at first order obtain the desired coupling terms
\be
\half \frac{M^4}{(M^2 + \half v^2)^2} \, g^2 v \, W_\mu^- W^{\mu +} h
+ \quart \frac{M^4(M^2 - \half v^2)}{(M^2 + \half v^2)^3} \,
(g^2 + g'^2) v \, Z_\mu Z^\mu h \,.
\ee
Now we convert to the rescaled quantities (\ref{rescaledvh}),
and expand to linear order in $\frac{v_{\rm SM}^2}{M^2}$ to obtain
\be
\half g^2 v_{\rm SM} \, W_\mu^- W^{\mu +} H
+ \quart (g^2 + g'^2) \left( 1 - \frac{v_{\rm SM}^2}{M^2} \right) v_{\rm SM}
\, Z_\mu Z^\mu H \,.
\ee
The couplings are therefore
\be
g_{WWH} = \half g^2 v_{\rm SM} \,, \quad
g_{ZZH} = \quart (g^2 + g'^2) \left( 1 - \frac{v_{\rm SM}^2}{M^2}
\right) v_{\rm SM} \,.
\ee

In the SM, these couplings are
\be
g_{WWH} = \half g^2 v_{\rm SM} \,, \quad
g_{ZZH} = \quart (g^2 + g'^2) v_{\rm SM} \,,
\ee
so the $W$-boson coupling is unchanged at this order in the $\CP^2$
SMEFT, but the $Z$-boson coupling is reduced by the factor
$1 - \frac{v_{\rm SM}^2}{M^2}$.

The current experimental constraints on these couplings from ATLAS and
CMS at the Large Hadron Collider are quite tight \cite{PDG}. The measurements
agree with the SM predictions to within 5\%. If, for the sake of
argument, one assumes the $Z$-boson coupling is less than 1\% below the SM
value, then $M$ would be constrained to be greater than 2.5 TeV.
  
%%%%%%%%%%%%%%%%%%%%%%%%%%%%%%%%%%%%%%%%%%%%%%%%%%%%%%%%%%%%%%%%%%%%%%%%%
\section{Conclusions}
%%%%%%%%%%%%%%%%%%%%%%%%%%%%%%%%%%%%%%%%%%%%%%%%%%%%%%%%%%%%%%%%%%%%%%%%%

We have proposed a highly constrained SMEFT, where the Higgs field
takes its value in the complex projective plane $\CP^2$ rather than in $\C^2$.
This is a truly nonlinear modification of the Higgs sector of the Standard
Model. The Lagrangian uses the Fubini--Study metric on $\CP^2$, and makes
some use of its $SU(3)$ symmetry. The only new parameter is a BSM mass scale
$M$, and the modification of the Standard Model is small provided $M \gg
v$, with $v$ the Higgs vev. The proposed model is fully consistent
with $U(2)$ gauge invariance; however, the custodial $SO(4)$
global symmetry of the Higgs sector of the Standard Model is mildly
broken, because the $U(2)$ orbits on $\CP^2$ are squashed
rather than round 3-spheres. The key prediction is that, as a
consequence of this squashed geometry, the $W$-boson mass is
closer to the $Z$-boson mass than in the Standard Model, while the
weak mixing angle $\theta_w$ remains unchanged. Yukawa couplings of
the Higgs boson to fermions are also modified by a universal factor.
Almost all current data are consistent with the Standard Model, so, as
for other BSM scenarios, there is no compelling evidence yet in
favour of the $\CP^2$ SMEFT. The data place a lower bound of O(7 TeV)
on $M$, unless the CDF measurement of the $W$-boson mass turns out
to be correct, in which case some physics beyond the Standard Model
would be needed to explain this, and within the $\CP^2$ SMEFT,
$M$ would be about 4 TeV. More precise and consistent data for the
gauge boson masses, Yukawa couplings and the Higgs boson self-couplings
would all be helpful.

The $\CP^2$ SMEFT has no obvious UV-completion. Possibly, it can be
extended to a UV-complete theory where the $SU(3)$ symmetry is
dynamically realised and spontaneously broken to $U(1)$ in two steps,
but that may require several additional gauge and Higgs fields, and additional
fermions too. The simplest idea would be to first break the $SU(3)$ symmetry
with a complex triplet Higgs field, but the unbroken gauge group would
then be $SU(2)$ rather than $U(2)$, and the projective space $\CP_2$
would not appear. There are several proposed unified models with
gauge group $SU(3)$ (or products of $SU(3)$ with another group) acting
linearly on multiple Higgs fields, e.g. \cite{Wei,Pug,SPV,Ng,Cza,HLW}.
These models mostly have substantially more gauge bosons, scalar
particles and fermions than the Standard Model, some with exotic
charges. These models do not seem to be equivalent to, or extensions of, the
$\CP^2$ SMEFT proposed here, although the gauge boson mass formulae in
\cite{Cza} are very similar to our formulae (\ref{gaugebosonmasses}).

\vspace{4mm}

%%%%%%%%% Acknowledgements %%%%%%%%%%%%%
\section*{Acknowledgements}
%%%%%%%%%%%%%%%%%%%%%%%%%%%%%%%%%%%%%%%%

I am grateful to Mia West and Eetu Loisa for stimulating discussions
about SMEFTs, initiated at the STFC HEP Forum: Completing the Higgs-saw
puzzle, Cosener's House, Abingdon, November '23. I also thank Alex Mitov for
comments, the anonymous referee for several constructive suggestions,
and finally Alessandra Cappati for recommending calculation of
the $WWH$ and $ZZH$ couplings during the IOP conference WIN 2025, Univ.
Sussex, June '25. This work has been partially supported by STFC
consolidated grant ST/P000681/1.

\end{document}